\documentclass[twocolumn,english,showpacs,floatfix,amsmath,amssymb,prl,
superscriptaddress]{revtex4}
\usepackage[T1]{fontenc}
\usepackage[latin9]{inputenc}
\usepackage{babel}

\usepackage{graphicx}
\usepackage{amssymb}
\usepackage{esint}
\usepackage{bm}



\begin{document}

\title{Scattering theory of current-induced forces in mesoscopic systems}

\author{Niels Bode} 
\affiliation{\mbox{Dahlem Center for Complex Quantum Systems and Fachbereich
Physik, Freie Universit\"at Berlin, 14195 Berlin, Germany}}

\author{Silvia Viola Kusminskiy} 
\affiliation{\mbox{Dahlem Center for Complex Quantum Systems and Fachbereich
Physik, Freie Universit\"at Berlin, 14195 Berlin, Germany}}

\author{Reinhold Egger} 
\affiliation{Institut f\"ur Theoretische Physik, Heinrich-Heine-Universit\"at,
D-40225 D\"usseldorf, Germany}

\author{Felix von Oppen}
\affiliation{\mbox{Dahlem Center for Complex Quantum Systems and Fachbereich
Physik, Freie Universit\"at Berlin, 14195 Berlin, Germany}}

\date{\today}
\begin{abstract}
We develop a scattering theory of current-induced forces exerted by the conduction electrons of a general mesoscopic conductor on slow ``mechanical'' degrees of freedom. Our theory describes the current-induced forces both in and out of equilibrium in terms of the scattering matrix of the phase-coherent conductor. Under general nonequilibrium conditions, the resulting mechanical Langevin dynamics is subject to both non-conservative and velocity-dependent Lorentz-like forces, in addition to (possibly negative) friction. We illustrate our results with a two-mode model inspired by hydrogen molecules in a break junction which exhibits limit-cycle dynamics of the mechanical modes. 
\end{abstract}
\maketitle

{\it Introduction.---}Current-induced forces play a central role in many contexts of modern condensed matter physics such as molecular electronics \cite{nitzan}, nanoelectromechanical systems (NEMS) \cite{nems}, spintronics \cite{spintorque}, or backaction forces in quantum measurements \cite{naik,stettenheim,bennett}. In its simplest incarnation, the current flow through, say, a carbon nanotube suspended over a gate electrode governs the electronic occupation of and thus the capacitive force exerted on the flexible nanotube. This force, including the dissipation of the mechanical motion resulting from retardation effects, was recently observed in seminal experiments \cite{adrian2, Steele09}. 
The nature of the interaction of conduction electrons with collective degrees of freedom, such as vibrational modes in NEMS or the magnetization in spintronics devices, depends sensitively on the relevant time scales. In the context of NEMS and molecular electronics, much work has focused on the limit in which the vibrational degree of freedom is fast compared to typical time scales of the conduction electrons. In this regime, the electron-vibron interaction causes vibrational sidebands in the current-voltage characteristics \cite{nitzan}, Franck-Condon blockade of transport \cite{ensslin}, or current-induced heating of the vibrational mode \cite{nitzan}. 

Here, we consider the opposite regime in which the collective mode is slow compared to electronic time scales. This limit applies for instance to the flexural modes of suspended graphene membranes and carbon nanotubes \cite{Steele09}, the current-induced electromigration of nano-scale islands on Ag surfaces \cite{tao}, the cooling and amplification of mechanical motion by the mesoscopic backaction of electrons \cite{naik,stettenheim}, certain molecular switches \cite{nitzan, martin1}, spin transfer torques and current-induced domain wall motion in ferromagnets \cite{spintorque}, as well as NEMS near continuous mechanical instabilities \cite{weick}. In this nonequilibrium Born-Oppenheimer (NEBO) limit, it is typically appropriate to describe the collective mode as a slow classical degree of freedom exhibiting Langevin dynamics. The central subject of this paper are the current-induced forces governing this Langevin process. Given the wide applicability and physical transparency of the scattering approach to quantum transport in nanoscale systems \cite{nazarov}, it seems highly desirable to develop a scattering theory of current-induced forces for a general phase-coherent mesoscopic conductor. Such a theory is the main result of the present paper. 

When developing a general theory of these current-induced forces out of equilibrium, it is imperative to consider more than one collective coordinate  \cite{dundas,lu}. Indeed, this introduces several qualitatively new and remarkable features into the Langevin equation
\begin{equation} 
\label{langevin}
M_\nu \ddot{X}_\nu + \frac{\partial U}{\partial X_\nu} = F_\nu - \sum_{\nu'} \gamma_{\nu\nu'} \dot{X}_{\nu'} +\xi_\nu
\end{equation}
for the collective coordinates $X_\nu$. The left-hand side describes the purely elastic dynamics and the right-hand side collects the current-induced forces, including the fluctuating Langevin force $\xi_\nu({\bf X})$. In the presence of an applied bias, the current-induced average force $F_\nu({\bf X})$ will generally be {\it non}-conservative. Moreover, the velocity-dependent force does not only contain a frictional contribution, corresponding to the symmetric part of the tensor $\gamma_{\nu\nu'}({\bf X})$, but also a Lorentz force (or Berry-phase contribution \cite{berry}), arising from its antisymmetric part. We will see below that these features can cause unconventional collective dynamics and may provide the operating principle of a nano-scale motor. 

{\it Model.---}Our starting point is a simple yet very general model of a mesoscopic conductor \cite{nazarov}. The ``quantum dot'' is described in terms of $M$ electronic orbitals with fermion operators $d_n$ obeying the Hamiltonian $H_D=\sum_{nn'} d_n^\dagger (h_D)_{nn'} d_{n'}$. The conduction electrons are coupled to a collection of $N$ ``slow'' collective degrees of freedom ${\bf X}(t)= (X_1,\ldots,X_N)$, which may, for instance, represent vibrational modes or atomic locations. We take this coupling to be linear, $h_D =  h_0 + \sum_{\nu=1}^N \Lambda_\nu \hat X_\nu$, characterized by the coupling matrices $\Lambda_\nu$. As usual in applications of scattering theory to mesoscopic transport, we assume that the electrons are otherwise non-interacting. In addition, the total Hamiltonian $H = H_D + H_L + H_T + H_X$ models the leads (labeled by $\alpha$ and with chemical potentials $\mu_\alpha$) as free Fermi systems, $H_L= \sum_\eta (\epsilon_{\eta}-\mu_\alpha) c_{\eta}^\dagger c^{}_{\eta}$ with composite index $\eta=({\bf k,\alpha})$, in terms of the lead fermions $c_\eta$ with Fermi-Dirac distribution $f_\alpha(\epsilon_\eta)$. It also includes hybridization between leads and quantum dot through $H_T = \sum_{\eta,n} c_\eta^\dagger W_{\eta n} d_n^{} + {\rm h.c.}$ with coupling matrix $W$, and describes the uncoupled mode dynamics through the Hamiltonian $H_X= \sum_\nu \hat P_\nu^2/2M_\nu + U({\bf X})$. 

{\it Non-equilibrium Born-Oppenheimer approximation.---}Generalizing previous formulations for $N=1$ \cite{martin1}, we start by deriving Green's function (GF) expressions for the current-induced forces in Eq.\ (\ref{langevin}) within the Keldysh formalism, based on the Heisenberg equations of motion 
\begin{equation}
   M_\nu \ddot X_\nu + \frac{\partial U}{\partial X_\nu} = - \sum_{n,n^\prime} d^\dagger_n (\Lambda_\nu)_{nn^\prime}d_{n^\prime}.
   \label{HEOM} 
\end{equation} 
(An alternative derivation proceeds in terms of Keldysh functional integrals, following e.g.\ Ref.~\cite{zazunov}.) In the NEBO approach, we average the force operator appearing on the right-hand side over time intervals long compared to the electronic time scales but short compared to time scales of the collective dynamics. In addition to a fluctuating force $\xi_\nu$ discussed below, this yields a time-averaged force ${\rm tr} [i \Lambda_\nu {\cal G}^<(t,t)]$. Here, the trace acts in dot level space, and the lesser GF is ${\cal G}_{nn'}^<(t,t')= i\langle d_{n'}^\dagger(t')d_n^{}(t)\rangle$. Note that this electronic GF should be evaluated for a given time dependence ${\bf X}(t)$ of the collective coordinates. For slow collective dynamics, to linear order in the velocities $\dot X_\nu$, we find (see below)
\begin{eqnarray}
  {\cal G}^< &\simeq& G^< + \frac{i}{2}\left( \partial_\epsilon G^< {\bf \Lambda} \cdot \dot {\bf X} G^A -  G^< {\bf \Lambda} \cdot \dot {\bf X} \partial_\epsilon G^A \right.
   \nonumber\\
   &&\left. +\partial_\epsilon G^R {\bf \Lambda} \cdot \dot {\bf X} G^< -  G^R {\bf \Lambda} \cdot \dot {\bf X} \partial_\epsilon G^< \right)
\label{NEBO}
\end{eqnarray}
in terms of the \textit{frozen} (adiabatic) GF $G(\epsilon,{\bf X})$ obtained for fixed configuration ${\bf X}$.
Inserting Eq.\ (\ref{NEBO}) into the time-averaged force given above, we obtain GF expressions for the current-induced forces in Eq.\ (\ref{langevin}). The mean force
\begin{equation}\label{mean}
F_\nu({\bf X}) = - \int \frac{d\epsilon}{2\pi i} \ {\rm tr} \left  [G^<(\epsilon,{\bf X}) \Lambda_\nu \right ]
\end{equation}
follows from the strictly adiabatic (frozen) limit ${\cal G}^< \simeq G^<$. The first non-adiabatic correction of ${\cal G}^<$ gives rise to the velocity-dependent term in Eq.~(\ref{langevin}). Decomposing $\gamma = \gamma^s + \gamma^a$ into symmetric and antisymmetric contributions, we obtain for the frictional contribution 
\begin{eqnarray} 
\label{dampmat}
\gamma^s_{\nu\nu'}({\bf X}) = \int \frac{d\epsilon}{2\pi} {\rm tr} \left\{ \Lambda_\nu G^<\Lambda_{\nu'} 
\partial_\epsilon G^> \right \}_s 
\end{eqnarray}
and for the effective orbital magnetic field 
\begin{eqnarray}
\gamma^a_{\nu\nu'}({\bf X}) = - \int\frac{d\epsilon}{2\pi} {\rm tr} \left\{ \Lambda_\nu G^<\Lambda_{\nu'}
\partial_\epsilon \left(G^A+G^R\right) \right\}_a. 
\label{effB}
\end{eqnarray}
Here, we employed $G^>=G^<+G^R -G^A$ with $G^A=(G^R)^\dagger$ and the shorthand $\{A_{\nu\nu'}\}_{s/a} = (A_{\nu\nu'}\pm A_{\nu'\nu})/2$.

The fluctuating force $\xi_\nu$ in Eq.\ \eqref{langevin} can be obtained from the (thermal and nonequilibrium) fluctuations of the force operator $- \sum_{n,n^\prime} d^\dagger_n (\Lambda_\nu)_{nn^\prime}d_{n^\prime}$ in Eq.\ (\ref{HEOM}). Exploiting Wick's theorem, one readily finds that the fluctuating force is a Gaussian white noise variable, $\langle \xi_\nu(t) \xi_{\nu'}(t')\rangle= D_{\nu\nu'}({\bf X}) \delta(t-t')$, with symmetric variance matrix
\begin{equation}\label{variance}
D_{\nu\nu'}({\bf X}) = \int\frac{d\epsilon}{2\pi} \ {\rm tr}\left
\{\Lambda_\nu G^< \Lambda_{\nu'} G^>\right\}_s.
\end{equation}
The correlations of $\xi_\nu(t)$ are local in time since in the NEBO limit, the electronic fluctuations involve much shorter time scales than the collective dynamics. For $N=M=1$, our expressions recover previous results \cite{martin1}. We emphasize that all current-induced forces in Eq.\ (\ref{langevin}) are nontrivial functions of the collective coordinates ${\bf X}$ as the latter enter into the frozen GF $G(\epsilon,{\bf X})$.
 
We briefly sketch the derivation of Eq.\ (\ref{NEBO}). Following Refs.\ \cite{martin1,pump}, we start from the Dyson equation ${\cal G}^R(i\partial_{t'}-H_D-\Sigma^R) = {\bf 1}$ for the retarded dot GF. The coupling between dot and leads is accounted for by the self-energy $\Sigma^R= - i \sum_{\alpha}  \Gamma_\alpha$ in terms of the hybridization matrices $\Gamma_\alpha= \pi \nu W^\dagger \Pi_\alpha W$ ($\nu$ is a lead density of states and $\Pi_\alpha$ a projection operator to states in lead $\alpha$). The Dyson equation can now be solved to linear order in $\dot{\bf X}$ by passing to the Wigner representation. Relating ${\cal G}^<$ to ${\cal G}^R$ using the Langreth rules yields Eq.\ (\ref{NEBO}).

{\it Scattering theory of current-induced forces.---}The scattering approach to quantum transport in mesoscopic systems, as epitomized by the Landauer-B\"uttiker formula for the conductance, is formulated in terms of the $S$-matrix describing the single-particle scattering by the nanostructure of electrons incident from one lead. For the model considered here, the frozen $S$-matrix is readily related to the frozen retarded GF $G^R(\epsilon,{\bf X})$ through \cite{nazarov}
\begin{equation}
  S(\epsilon,{\bf X}) = 1 - 2\pi i \nu W G^R(\epsilon,{\bf X}) W^\dagger .
  \label{SMatrix}
\end{equation}
We assume the wide-band limit in which $\nu$ and $W$ are energy independent.
The average force $F_\nu({\bf X})$ in Eq.\ (\ref{mean}) can be written in terms of the frozen $S$-matrix [Eq.\ (\ref{SMatrix})] by expressing the lesser GF $G^<=G^R\Sigma^< G^A$ in terms of the self-energy $\Sigma^<(\epsilon)=2 i \sum_{\alpha} f_\alpha \Gamma_\alpha$ \cite{nazarov}. Using the identity $2\pi i \nu W^\dagger W = (G^R)^{-1}-(G^A)^{-1}$, we then find 
\begin{equation} \label{mean2}
 F_\nu({\bf X}) = \sum_\alpha \int \frac{d\epsilon}{2\pi i}  f_\alpha \
{\rm Tr} \left(\Pi_\alpha S^\dagger \frac{\partial S}{\partial X_\nu} \right)
\end{equation}
for the mean force expressed through the $S$-matrix. Here the trace ``Tr'' acts in lead-channel space. By a similar set of steps, we can also express the variance [Eq.~(\ref{variance})] of the fluctuating force in terms of the frozen $S$-matrix,
\begin{eqnarray}
\label{variance2}
D_{\nu\nu'}({\bf X})&=&  \sum_{\alpha\alpha'} \int\frac{d\epsilon}{2\pi}
f_\alpha(1-f_{\alpha'}) \\ \nonumber
 &\times& {\rm Tr} \left\{ \Pi_\alpha \left( S^\dagger 
\frac{\partial S}{\partial X_{\nu}}\right)^\dagger \Pi_{\alpha'} S^\dagger 
\frac{\partial S}{\partial X_{\nu'}} \right\}_s.
\end{eqnarray}
By going to a basis in which $D$ is diagonal and using $\Pi_\alpha=\Pi_\alpha^2$, we find that $D$ is a positive definite matrix. 

To express the velocity-dependent forces in terms of the scattering matrix in general nonequilibrium situations, we need to go beyond the frozen scattering matrix $S$. The relevant concepts have been developed in the context of the scattering theory of quantum pumps \cite{pump}. For adiabatic parameter variations, the full ${\cal S}$-matrix of mesoscopic conductors can be expressed in Wigner representation as ${\cal S}(\epsilon,t)= 1-2\pi i \nu W {\cal G}^R(\epsilon,t) W^\dagger$, which naturally extends Eq.\ (\ref{SMatrix}). Expanding ${\cal S}$ to linear order in the velocities $\dot{\bf X}$ of the adiabatic variables, Moskalets and B\"uttiker \cite{pump} introduced an $A$-matrix through ${\cal S}(\epsilon,t) \simeq S(\epsilon,{\bf X}(t))+ \sum_\nu \dot X_\nu(t) A_\nu(\epsilon,{\bf X}(t))$, where
\begin{equation} \label{Amat}
A_\nu(\epsilon,{\bf X}) = \pi \nu W ( \partial_\epsilon G^R \Lambda_\nu G^R - 
G^R \Lambda_\nu \partial_\epsilon G^R ) W^\dagger
\end{equation}
in terms of the frozen GF.  It is now tedious but straightforward to write the damping coefficients in Eq.~(\ref{dampmat}) as 
\begin{eqnarray}\nonumber
&&\gamma^s_{\nu\nu'} = \sum_\alpha \int \frac{d\epsilon}{4\pi} (-\partial_\epsilon
f_\alpha) {\rm Tr} \left\{ \Pi_\alpha \frac{\partial S^\dagger} {\partial X_\nu} \frac{\partial S}{\partial X_{\nu'}}\right\}_s \\ 
 \label{damp2}
 &&+ \sum_{\alpha} \int \frac{d\epsilon}{2\pi i} f_\alpha
{\rm Tr} \left\{ \Pi_{\alpha}\left( \frac{\partial S^\dagger}{\partial X_\nu} A_{\nu^\prime} - A^\dagger_{\nu^\prime} \frac{\partial S}{\partial X_\nu} \right) \right\}_s .
\end{eqnarray}
Similarly, the effective magnetic field in Eq.~\eqref{effB} becomes
\begin{equation}
\gamma^a_{\nu\nu^\prime}= \sum_{\alpha}\int\frac{d\epsilon}{2\pi i} f_\alpha
{\rm Tr}\left\{\Pi_{\alpha}\left(S^\dagger \frac{\partial A_{\nu}}{\partial X_{\nu'}}-\frac{\partial A^\dagger_{\nu}}{\partial X_{\nu'}} S\right)\right \}_a.
\label{effB2}
\end{equation}
Several comments on these results are now in order.

(i) The current-induced mean force $F_\nu({\bf X})$ in Eq.~\eqref{mean2} is conservative  (the gradient of a potential) if its ``curl''
\begin{eqnarray}\label{curl}
\frac{\partial F_\nu}{\partial X_{\nu'}}-\frac{\partial F_{\nu'}}
{\partial X_\nu} = \sum_\alpha \int \frac{d\epsilon}{\pi i}
f_\alpha  {\rm Tr} \left( \Pi_\alpha \frac{\partial S^\dagger}{\partial X_\nu}
\frac{\partial S}{\partial X_{\nu'}} \right)_a
\end{eqnarray}
vanishes. This is indeed the case in thermal equilibrium, where Eq.~\eqref{curl} can be turned into a trace over a commutator of finite-dimensional matrices due to the relations $f_\alpha=f$, $\sum_\alpha \Pi_\alpha=1$, and unitarity $S^\dagger S={\bf 1}$ implying $\partial (S^\dagger S)/\partial X_\nu =0$. In general, however, the mean force will be {\it non}-conservative in out-of-equilibrium situations. Equation \eqref{mean2} for ${\bf F}({\bf X})$ provides an $S$-matrix representation for the ``water wheel'' \cite{dundas} or ``electron wind'' \cite{lu} force discussed recently.

(ii) The second contribution to the damping matrix (\ref{damp2}) is a pure nonequilibrium term. In fact, it can be shown to vanish in equilibrum using unitarity of $S$ as well as ${\cal S}$, where the latter implies \cite{pump} ${\bf A}S^\dagger + S {\bf A}^\dagger = (i/2)[\partial_{\bf X} S\partial_\epsilon S^\dagger - \partial_\epsilon S \partial_{\bf X} S^\dagger]$. The first term in Eq.~(\ref{damp2}) is a close analog of the $S$-matrix expression for Gilbert damping in ferromagnets \cite{gilbert}. From this perspective, our expressions for $\gamma^s$ can be viewed as mechanical analogs of Gilbert damping, generalized to arbitrary nonequilibrium situations. While the first term has strictly positive eigenvalues, the second term can have either sign and even cause negative eigenvalues of the full dissipation matrix $\gamma^s$.  (For a simple model exhibiting negative damping, cf.~Ref.~\cite{brandes}.)  This reflects the fact that out of equilibrium, the electronic degrees of freedom can effectively pump energy into the collective modes. 

(iii) If the conductor is time-reversal symmetric such that $S=S^T$ and $A=-A^T$, the effective orbital magnetic field $\gamma^a$ vanishes in thermal equilibrium. In general out-of-equilibrum situations, $\gamma^a$ is nonzero even for time-reversal symmetric conductors since the current provides another quantity which is odd under time reversal.

(iv) In thermal equilibrium, the variance $D_{\nu\nu'}$ of the fluctuating force
is related to the dissipation matrix $\gamma^s$ through the
fluctuation-dissipation theorem $D_{\nu\nu'} = 2 T \gamma^s_{\nu\nu'}$. This
follows readily using that the second term in Eq.~\eqref{damp2} vanishes in
equilibrium and that $-\partial_\epsilon f= f(1-f)/T$. We note that a
recent study \cite{bennett} of backaction forces in quantum measurements in
terms of the $S$-matrix obtains a friction coefficient which violates the fluctuation-dissipation
theorem unless the S-matrix is energy independent.

(v) The current flowing in lead $\alpha$ is given in terms of an appropriate average over the Langevin process of the current to linear order in $\dot{\bf X}$. Starting with the GF expression, this is the sum of the conventional Landauer-B\"uttiker current \cite{nazarov}, $I_\alpha({\bf X})=(e/h)\sum_{\beta}\int d\epsilon (f_\alpha-f_\beta) \ {\rm Tr} (S\Pi_\beta S^\dagger \Pi_\alpha)$, and a pumping correction \cite{pump}, 
\begin{eqnarray}
&&\delta I_\alpha ({\bf X}) =  \frac{e}{2\pi} \int d\epsilon \sum_\beta
 \dot{\bf X}\cdot \left[  -\frac{\partial f_\beta}{\partial \epsilon} 
 {\rm Im}{\rm Tr} \left( \Pi_\alpha \frac{\partial S}{\partial {\bf X}} \Pi_\beta S^\dagger \right) \right.  \nonumber\\
&& \left. +  f_\beta \ {\rm Re}{\rm Tr} \left( i\Pi_\alpha \frac{\partial S}{\partial{\bf X}} \Pi_\beta \frac{\partial S^\dagger}{\partial \epsilon} - 2\Pi_\alpha {\bf A} \Pi_\beta S^\dagger  \right) \right] .
\label{pump}
\end{eqnarray}
In thermal equilibrium, the first term reproduces the well-know $S$-matrix expression \cite{brouwer} for the pumping current, while the second term vanishes by unitarity of ${\cal S}$. 

{\it Limit-cycle dynamics.---}We now employ our general theory to show that at finite bias, the current-induced forces can change the collective dynamics qualitatively. Without applied bias, the collective dynamics exhibits oscillations about a stable equilibrium. These oscillations can become destabilized under out-of-quilibrium conditions, due to negative eigenvalues of the dissipation matrix $\gamma^s$, a non-conservative force field ${\bf F}({\bf X})$, or both. Within a stability analysis for these scenarios, one would predict an exponential instability of at least one collective mode \cite{lu}.

\begin{figure*}[t]
\includegraphics[width=4
cm,keepaspectratio=true]{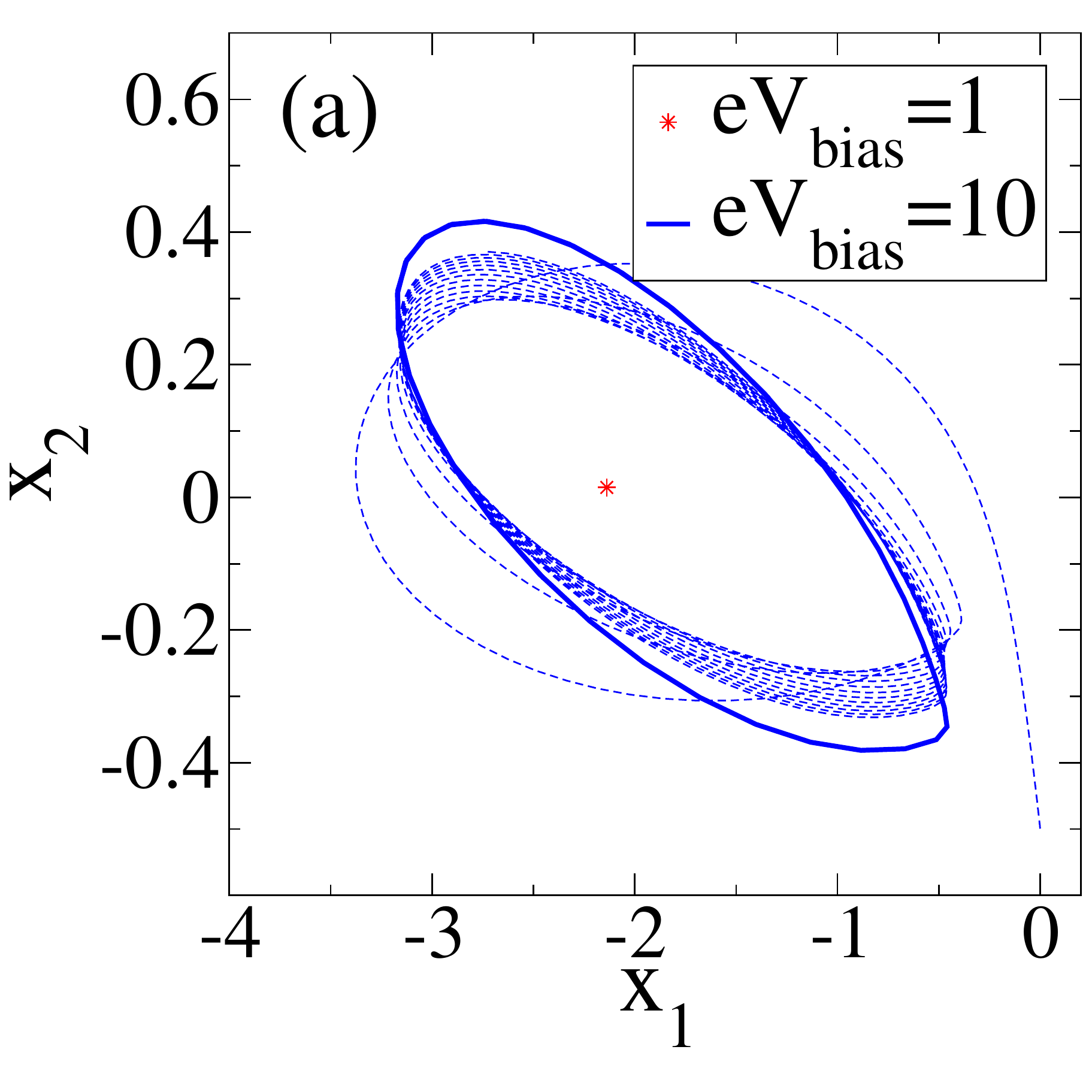}\,\,\,\,\includegraphics[width=4
cm,keepaspectratio=true]{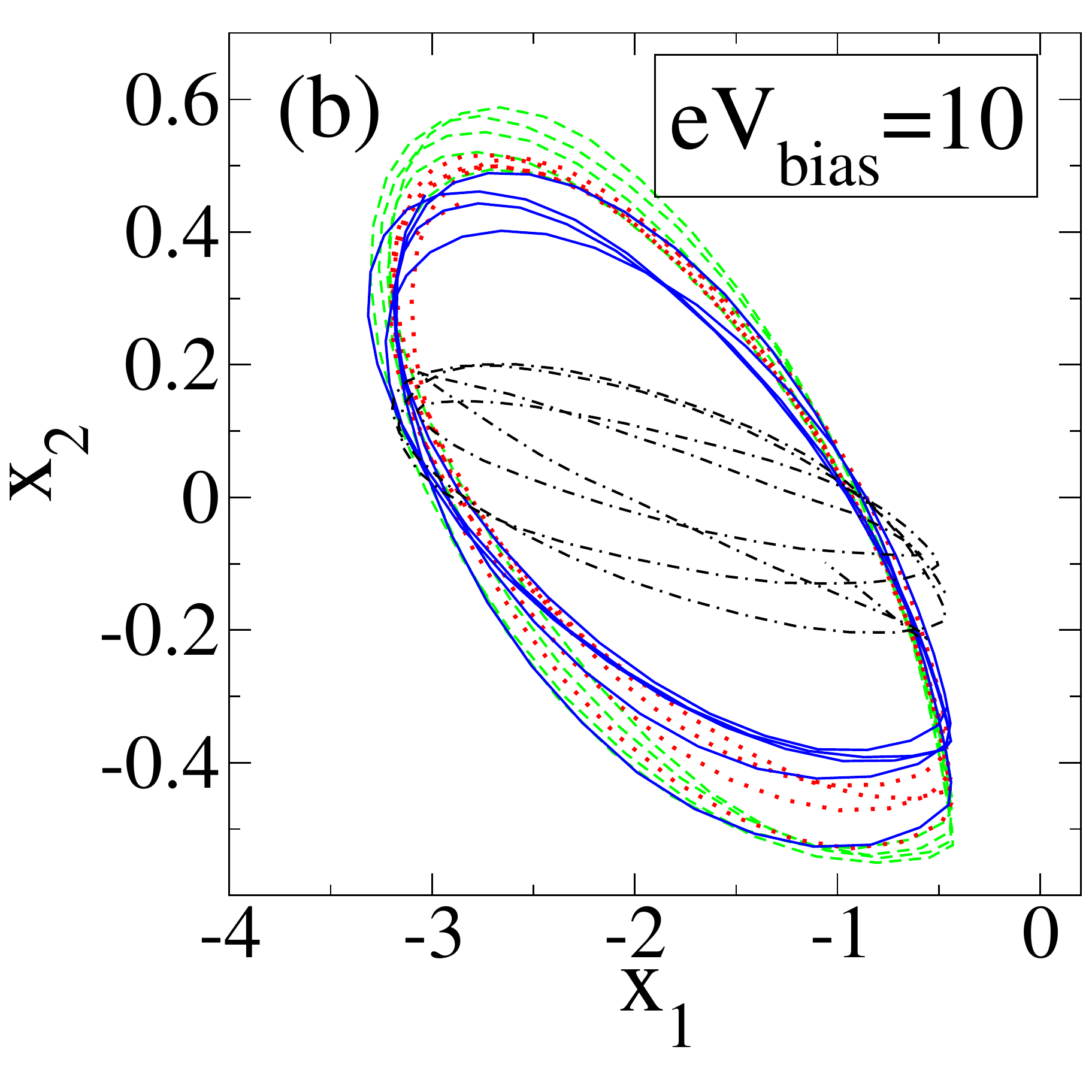}\,\,\,\,\includegraphics[width=4
cm,keepaspectratio=true]{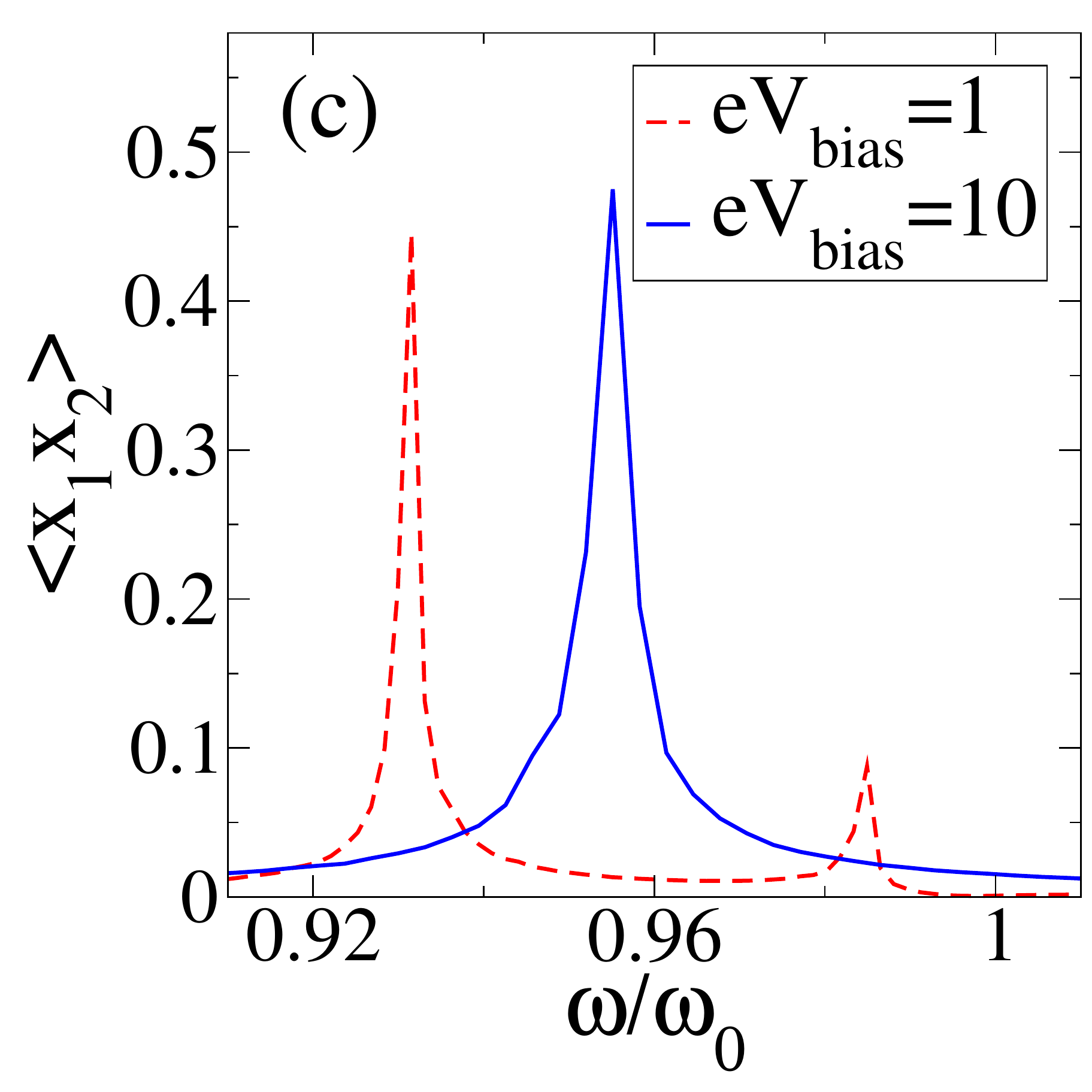}\,\,\,\,\includegraphics[width=4
cm,keepaspectratio=true]{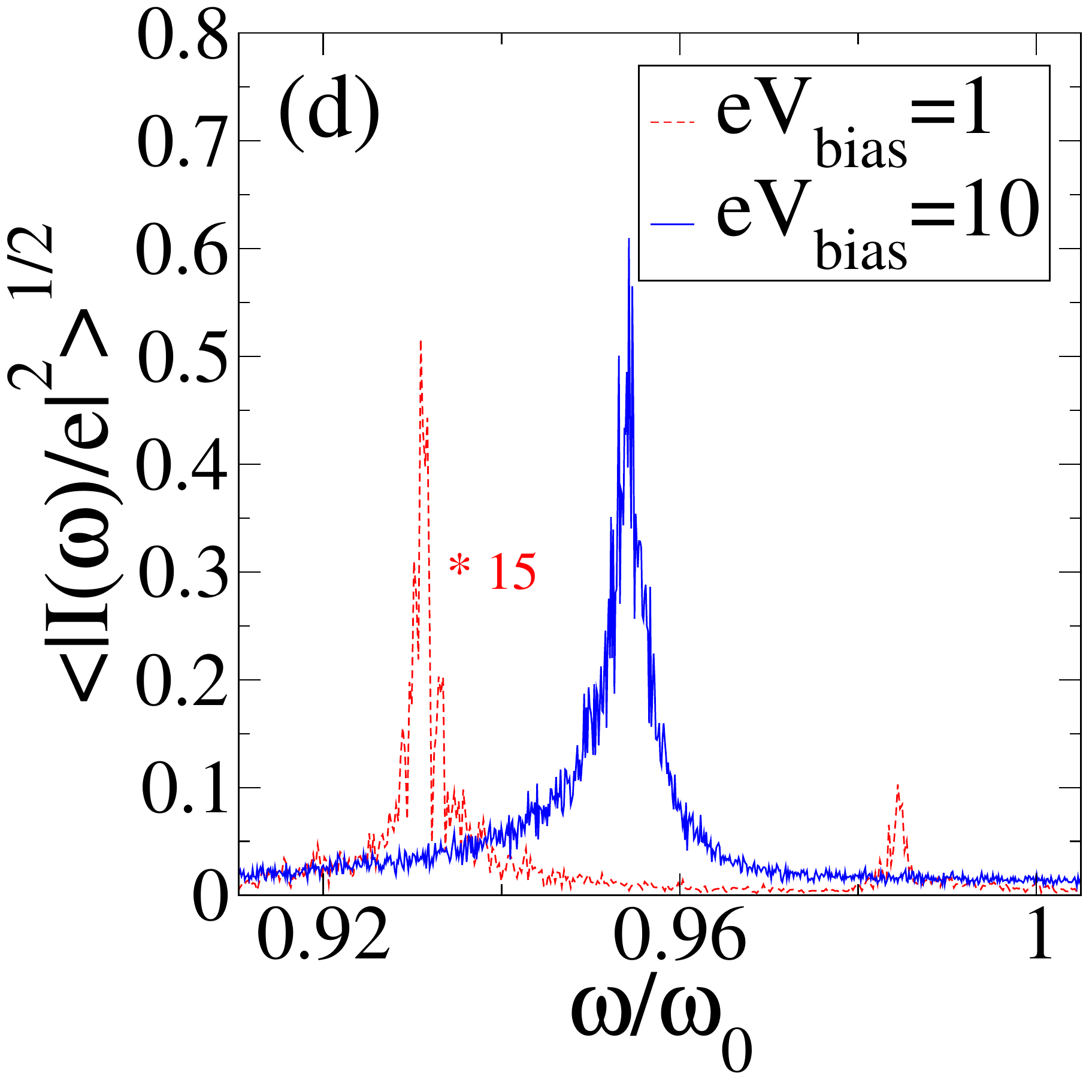}
\caption{(Color online) (a) Limit cycle (blue solid line) and its approach (blue dotted line) at large bias vs stable oscillations at low bias (red asterisk) in the Langevin dynamics without fluctuating force. (b) Several periods of typical trajectories with fluctuating forces for the parameters of the limit cycle in (a). (c) Fourier transform of the correlation function $\langle X_1(t) X_2(t+\tau)\rangle$. The limit cycle is signaled by a {\em single} peak, as opposed to {\em two} peaks in the absence of a limit cycle. (d) The same signature appears in the current-current correlation function, making the onset of limit-cycle dynamics observable in experiment. These zero-temperature results, based on analytical results for the current-induced forces, are obtained for $\Lambda_1 = \lambda_1 \sigma_0$ and $\Lambda_2 = \lambda_2 \sigma_x$ with $\lambda_1/\lambda_2=3/2$. In units of  ${\lambda^2}/{(M\omega_0^2)}$ [where $\lambda=({\lambda_1+\lambda_2})/{2}$], the elastic modes are degenerate with $\hbar \omega_0 = 0.014$, $\Gamma_{L,R}=\frac{1\pm0.8}{2} (\sigma_0 \pm \sigma_z)$, and the hopping between the orbitals is $w=0.9$. The dimensionless coordinates are $x_i = ({M \omega_0^2}/{\lambda}) X_i$.}
\end{figure*}

A more thorough understanding of the resulting dynamics requires one to retain the full anharmonic Langevin dynamics, including the complete dependence of the current-induced forces on the mode coordinates ${\bf X}$, which is readily possible within our formalism. Indeed, we find that the current-induced instability typically drives the collective degrees of freedom into a limit cycle. Systems with more than one collective degree of freedom are most susceptible to this limit-cycle scenario when the elastic contributions to the mode frequencies are (near)degenerate.

Consider a model, inspired by transport measurements on hydrogen molecules between electrodes \cite{smit}, with two electronic orbitals which are each coupled to one of the leads and which allow tunneling between them. The hydrogen molecule in a single-molecule contact has two almost degenerate low-energy vibrational modes \cite{smit}, corresponding to the center-of-mass vibration between the contacts and the rigid-rotation mode. Taking their respective couplings $\Lambda_\nu$ to the electronic orbitals as Pauli matrices in  the electronic subspace, say $\sigma_0$ and $\sigma_x$, we find that the mean force becomes non-conservative in the presence of a finite bias. Without fluctuating force, this gives rise to limit-cycle dynamics at larger biases which evolves from the stable equilibrium at small voltages, see Fig.\ 1(a). With Langevin force [Fig.\ 1(b)], the trajectory fluctuates about the limit cycle, with occasional larger excursions.

Most importantly, this transition into limit-cycle dynamics has directly observable consequences even for the full Langevin dynamics. In the absence of a limit cycle, the collective motion is a random superposition of two normal modes whose frequencies differ due to the current-induced forces. (Here, we assume for simplicity that the elastic contributions to the mode frequencies are strictly degenerate.) Apart from higher harmonics, this leads to {\em two} distinct peaks in the Fourier transforms of correlation functions such as $\langle X_1(t)X_2(t+\tau)\rangle$. In contrast, once the limit cycle develops, its frequency completely dominates the correlation function, and its Fourier transform exhibits a single peak only [Fig.\ 1(c)]. This frequency locking is mirrored in the correlation function of the current through the NEMS, as shown in Fig.\ 1(d). This should serve as a direct signature of current-induced limit-cycle dynamics in experiment. Remarkably, the present system resembles a nano-scale motor as the mechanical motion remains approximately periodic even for the full Langevin dynamics and this periodic motion is driven by a non-conservative force. 

We acknowledge discussions with P. Brouwer, G. Zarand, and L. Arrachea as well as support by the DFG through SPP 1459, SFB TR/12 and SFB 658.


\begin{thebibliography}{12}

\bibitem{nitzan}
M. Galperin, M.A. Ratner, and A. Nitzan, J. Phys.: Cond. Matt. {\bf 19}, 103201 (2007).

\bibitem{nems} H.G. Craighead, Science {\bf 290}, 1532 (2000); M.L. Roukes,
Phys. World {\bf 14}, 25 (2001).

\bibitem{spintorque} D.C.\ Ralph and M.D.\ Stiles, J. Magn. Magn. Mater. {\bf 320}, 1190 (2008).


\bibitem{naik}
A. Naik \textit{et al.}, Nature {\bf 443}, 193 (2006).

\bibitem{stettenheim}
J. Stettenheim \textit{et al.}, Nature {\bf 466}, 86 (2010).

\bibitem{bennett} 
S.D. Bennett, J. Maassen, and A.A. Clerk, Phys. Rev. Lett. {\bf 105}, 217206 (2010).


\bibitem{Steele09} 
G.A. Steele \textit{et al.},
Science {\bf 325}, 1103 (2009).

\bibitem{adrian2}
B. Lassagne \textit{et al.}, 
Science {\bf 325}, 1107 (2009).


\bibitem{ensslin}
R. Leturcq \textit{et al.}, Nature Phys. {\bf 5}, 327 (2009).

\bibitem{tao}
C. Tao, W.G. Cullen, and E.D. Williams, Science {\bf 328}, 736 (2010).

\bibitem{martin1} 
D. Mozyrsky, M.B. Hastings, and I. Martin, Phys. Rev. B {\bf 73}, 035104 
(2006); F. Pistolesi, Ya.M. Blanter, and I. Martin, \textit{ibid.} {\bf 78}, 
085127 (2008).

\bibitem{weick} 
G. Weick \textit{et al.}, 
Phys. Rev. B {\bf 81}, 121409(R) (2010); G. Weick, F. von Oppen, and F. Pistolesi, {\it ibid.} {\bf 83}, 035420 (2011).

\bibitem{nazarov}
Yu.V. Nazarov and Ya.M. Blanter, \textit{Quantum transport} (Cambridge University Press, 2010).

\bibitem{dundas} 
T.N. Todorov, D. Dundas, and E.J. McEniry, Phys. Rev. B {\bf 81}, 075416 (2010).

\bibitem{lu} 
J.T. L\"u, M. Brandbyge, and P. Hedegard, Nano Lett. {\bf 10}, 1657 (2010). 

\bibitem{berry}
M.V. Berry and J.M. Robbins, Proc. R. Soc. London, Ser. A {\bf 442}, 659 (1993).

\bibitem{zazunov} A. Zazunov and R. Egger, Phys. Rev. B {\bf 81}, 104508 (2010).

\bibitem{pump}
M. Moskalets and M. B\"uttiker, Phys. Rev. B {\bf 69}, 205316 (2004);
L. Arrachea and M. Moskalets, {\it ibid.} {\bf 74}, 245322 (2006).

\bibitem{gilbert}
A. Brataas, Y. Tserkovnyak, and G.E.W. Bauer, Phys. Rev. Lett.
{\bf 101}, 037207 (2008).

\bibitem{brandes}
R. Hussein \textit{et al.}, 
Phys. Rev. B {\bf 82}, 165406 (2010).

\bibitem{brouwer} 
P.W. Brouwer, Phys. Rev. B {\bf 58}, R10135 (1998).

\bibitem{smit}
R.H.M. Smit \textit{et al.}, 
Nature {\bf 419}, 906 (2002).

\end{thebibliography}
\end{document}